# The Micro-Paper: Towards cheaper, citable research ideas and conversations

FRANK ELAVSKY, Carnegie Mellon University, fje@cmu.edu

Academic, peer-reviewed "short" papers are a common way to present a late-breaking work to the academic community that outlines preliminary findings, research ideas, and novel conversations. By comparison, blogging or writing posts on social media are an unstructured and open way to discuss ideas and start new conversations. Both have limitations in the proliferation of research ideas. The short paper format relies on the conference and journal submission process while blogging does not operate within a structured format or set of expectations at all. However, at times the demand exists for late-breaking ideas and conversations to arise in a raw form or with urgency but should still be archived and recorded in a way that promotes citational honesty and integrity. To address this, I present: The Micro-Paper, as a micro-paper itself. The Micro-Paper is a small, cheap, accessible, digital document that is self-published and archived, akin to a pre-print of a short paper. This meta micro-paper discusses the context, goals, and considerations of micro-paper authoring.

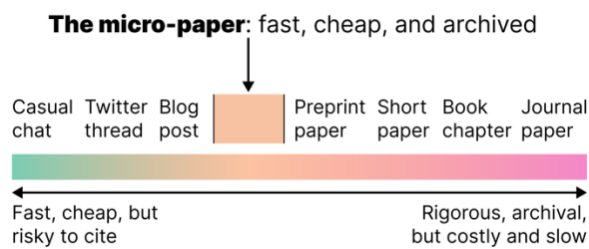

**Fig 1:** The micro-paper fills a gap on the spectrum between fast, cheap ideas and rigorous, archival work.

## 1. What is a micro-paper?
A micro-paper is a paper between 1 and 4 pages in length that engages a single idea clearly. A micro-paper can be anything from focused blog post to a preprinted short paper, but it is published through an open archive.

## 2. The context of micro-papers
Ideas and conversations often arise in research settings in response to problems, gaps, or issues. But some ideas and conversations also come about in a more generative fashion: they are still responding to something within a context but are not concerned with problems and gaps. So, whether filling gaps or otherwise, a micro-paper (and even the authors at times) must clearly and reflexively be situated within a conversational context.

At a meta-level, this paper *is* framed as a gap-filler (see: Fig 1). Currently, academic publishing is expensive, time-consuming, and high risk. And while much of the process could be argued as a necessary set of procedures to ensure we aren't making claims that are misleading or unfounded, there exists a gap for disseminating ideas that should be cheap, accessible, and intended to inspire other work [4]. Not all research conversation needs to have answers. The heart of research as a community is due to the dissemination of ideas that aren't only congealed or refined, but raw and messy as well.

Generally, the gap for disseminating cheap ideas is filled by academics today through Twitter, blogging, or in some form of social media or another. Sometimes the gap is filled through workshops, position papers, or conversation pieces (such as ACM's Interactions). There are many options for sharing ideas, all with different tradeoffs between editorial and authoring expenses, time, archiving, accessibility, and democratization of the process.

## 3. A micro-paper's goal is the free, cheap, open, and honest dissemination of ideas
The micro-paper's focus is on *ideas* for the sake of generative work, conversation, and inspiration. In contrast, a micro-paper is not an appropriate venue for sharing findings, claims, or experiments. The nature of methodological generation of knowledge is most trustworthy when there is a more rigorous process in place. Some avenues generate *good* or *trustworthy* knowledge and ideas, but the micro-paper is a place for sharing *potentially useful* ideas. Good or trustworthy knowledge may require more careful review [3, 4, 6], but potentially useful ideas should at least be archived.

### 3.1 A micro-paper must be small and cheap
Whether having peer-reviewed work or an editorial team, most writing is costly in both time and money. It is also risky: your work could get rejected or require slow iterations of feedback and review.

In contrast, this micro-paper took me 3 hours on a random Thursday in February (when I should be crunching for another deadline). Most micro-papers should ideally be short enough in length to encourage both rapid authoring and reading.

### 3.2 A micro-paper must be archived

While blogging and tweeting is cheap and fast and encourages ideas to be shared, these aren't trustworthy archives. And sometimes good ideas arise in these faster, cheaper contexts that should be captured, articulated, and stored for later reference. With the existential threat of twitter disintegrating at any moment and the entirely unmaintained space of many academic blogs, it is important for some ideas to be archived [5].

While some blog maintainers may have higher standards for their longevity, url stability, dois, and records of changes, it may make sense for most to use an existing pre-print archival platform, like arXiv for ensuring trustworthiness and reliability in your readers.

### 3.3 Revisions must also be archived

In addition to archival processes for the sake of accessing later, many archival sites (such as arXiv) also keep track of revision histories. The ephemeral and non-standardized way that individuals operate their own blogs and social media means that not only might something move or cease to exist (a findability problem) but there is also an honesty problem when contents change or update without record [7].

### 3.4 A micro-paper must be accessible

Blogs, twitter, and mastodon have done more for disability discourse than most other media, but especially more than the dreaded PDF favored by academics. This is because text-based media online (generally in HTML or Markdown) has immense accessibility potential over PDF [1] (and paper publications). All artifacts of academic discourse, including every paper publication, should be more accessible. Due to a micro-paper's size, it is easier for authors to learn accessibility for than a full-size paper with proprietary formatting involved. This micro-paper was authored in Microsoft Word, exported as a PDF, and then converted into HTML using pandoc.

With the push for arXiv to transition more towards accessible formats of publication [2], I believe that trustworthy archives that are accessibility-first are near. Micro-papers will compliment this push.

## 4. Considering when a micro-paper is the write choice (pun intended)

If someone has an idea, conversation, or late-breaking work, they might consider the following questions:

*Why wouldn't I write a short academic paper and submit to a traditional venue?*
It is time-consuming, expensive, and requires waiting for the publication cycle. It is also higher risk, in cases where the peer review process might reject it.

*Why shouldn't I write a piece in a non-peer reviewed publication, like ACM Interactions?*
This is also time-consuming and higher risk, because editorial interest may conflict. In addition, these often require an existing network of colleagues, invitation to contribute, or formal submission and selection process.

*Why shouldn't I write a blog or twitter thread?*
Blogs and social media posts raise concerns about archival quality and trustworthiness. Accessing the piece later may become difficult or cumbersome. Some great ideas and discussions have been lost in time due to the ephemeral nature of these cheap and fast options.

*When is writing a micro-paper a good idea?*
Notably, there is no formal peer review for a micro-paper. Our currently imagined peer review process may not make sense for all work published with the intent to push new ideas and conversations. It is even worth considering if this practice should continue at all [3, 4].

A practical use for a micro-paper may be as a preprint or early draft for an eventual short or full paper submission, position paper, or book chapter. The greatest strength of both the pre-print and short paper process is that they can invigorate scholars with new or raw ideas to see those turn into full projects. Short papers also have a core readership and opportunities to present that are not afforded to micro-papers, so for early career researchers it may be important to consider ways to use these two formats together.

For folks who simply want to get a citable idea out into the world without regard for submission and publication cycles and procedures, a micro-paper is a good choice as well.

And lastly, there may be authors with too many ideas to pursue (even when some are useful) and they are willing to admit that they won't pursue every idea that they have. A micro-paper is a way to put the idea into the discussion and let it run its course. In my case, I recognize that some problems and patterns are outside of the scope of my leverage and experience to address, such as contributions to design or behavioral domains of accessibility (when my area is strictly technical contributions).

## 5. Conclusion

The hope is that both the procedural and systemic inaccessibility of the short paper authoring process and the citational uncertainty of blogs and social media can be addressed with the micro-paper. I hope to see early, usable ideas shared more freely and especially hope to invigorate young scholars and include historically excluded folks, such as those with disabilities, in the larger research conversation.

## Acknowledgements
Special thanks to Jonathan Zong, for your encouragement to make my cheap ideas citable and for feedback on this micro-paper. Also thanks to the folks at the MIT Vis Lab (Arvind, Crystal, and Alan) for supporting my not-traditionally-publishable ideas over the past couple years especially. Hearing that my "tweets are a public service" encouraged me to make the heart of that service last longer than Twitter does (hopefully).